\documentclass[conference]{IEEEtran}
\IEEEoverridecommandlockouts
\usepackage{cite}
\usepackage{url}
\usepackage[caption=false, font=footnotesize]{subfig}
\usepackage{amsmath,amssymb,amsfonts}
\usepackage{algorithmic}
\usepackage{graphicx}
\usepackage{textcomp}
\usepackage{xcolor}
\def\BibTeX{{\rm B\kern-.05em{\sc i\kern-.025em b}\kern-.08em
    T\kern-.1667em\lower.7ex\hbox{E}\kern-.125emX}}
\begin{document}

\title{Advanced Gaze Analytics Dashboard\\
}

\author{\IEEEauthorblockN{Gavindya Jayawardena, Vikas Ashok, Sampath Jayarathna}
\IEEEauthorblockA{\textit{Department of Computer Science}, 
\textit{Old Dominion University}, Norfolk, VA, USA \\
gavindya@cs.odu.edu, vganjigu@cs.odu.edu, sampath@cs.odu.edu}
}

\maketitle

\begin{abstract}
Eye movements can provide informative cues to understand human visual scan/search behavior and cognitive load during varying tasks. Visualizations of real-time gaze measures during tasks, provide an understanding of human behavior as the experiment is being conducted. Even though existing eye tracking analysis tools provide calculation and visualization of eye-tracking data, none of them support real-time visualizations of advanced gaze measures, such as ambient or focal processing, or eye-tracked measures of cognitive load. In this paper, we present an eye movements analytics dashboard that enables visualizations of various gaze measures, fixations, saccades, cognitive load, ambient-focal attention, and gaze transitions analysis by extracting eye movements from participants utilizing common off-the-shelf eye trackers. We validate the proposed eye movement visualizations by using two publicly available eye-tracking datasets. We showcase that, the proposed dashboard could be utilized to visualize advanced eye movement measures generated using multiple data sources.

\end{abstract}


\section{Introduction}

The ``Eye-Mind Hypothesis''~\cite{just1980theory} suggests that the mind and eyes work together to form humans' perceptions of the world. 
The eyes and the mind are constantly adjusting and updating the visual information to create an accurate representation of the environment. 
Eyes move rapidly to keep the image of the world stable on the retina when people are moving, thus allowing the mind to construct a stable representation of the world.
Inspired by this real-time behavior of eyes and mind, real-time eye movement analysis can provide insights into various aspects of human behavior and cognition. 
For instance, in neuroscience research \cite{popa2015reading}, eye movements have proven to reveal information about how people perceive and process information, along with how their cognition varies with the tasks they are working on. This helps researchers better understand the workings of the brain and incorporate conditions like Attention-deficit/hyperactivity disorder (ADHD) \cite{jayawardena2020pilot}, Autism Spectrum Disorder (ASD), and concussion in activities.

Real-time visualizations of eye movement analytics can also potentially enhance safety, training, research, and user experiences across various domains. 
They enable users to make informed decisions, optimize their actions, and gain a deeper understanding of how attention and visual behavior impact their activities.
In applications like healthcare, real-time visualizations of eye movements can help medical professionals diagnose and treat patients more effectively. 
Another example 
is, monitoring driver's attention \cite{ahlstrom2021eye} and alertness in real time using eye movement data, to provide warnings or take corrective actions to prevent accidents.

Real-time visualizations of advanced gaze measures are also valuable for training and skill development. They allow trainees to see when and where their attention is focused and how it relates to their performance. 
Researchers in various fields, such as psychology, neuroscience, and human-computer interaction, can benefit from real-time visualizations to study cognitive processes, attention, and decision-making, as they provide immediate insights into subjects' behavior.

This study aims to address the existing research gap by
designing an eye movements analytics dashboard that provides visualizations of advanced gaze measures in real time. 
The proposed advanced gaze measures analytics dashboard provides visualizations of advanced gaze measures including ambient/focal attention coefficient $\mathcal{K}$~\cite{krejtz2016discerning}, an eye tracked measure of cognitive load, Real-time Index of Pupillary Activity (RIPA)~\cite{jayawardena2022toward},  and gaze transition matrices~\cite{krejtz2015-gaze-transition-entropy}, during the course of a single scan path of a participant in real-time. 
To validate the proposed dashboard in terms of the applicability of multiple data sources, we examine two scenarios drawn from two publicly available eye-tracking datasets and evaluate the advanced gaze measures visualizations.

In summary, our contributions in this work are:
\begin{enumerate}
  \item We design and develop an interactive, advanced gaze measures analytics dashboard that provides visualizations of positional gaze measures, as well as advanced gaze measures in real-time.
  \item We demonstrate the utility of the proposed dashboard by examining two scenarios drawn from two publicly available eye-tracking datasets.
\end{enumerate}

\begin{figure}[h]
\centering
   \frame{\includegraphics[width=0.48\textwidth]{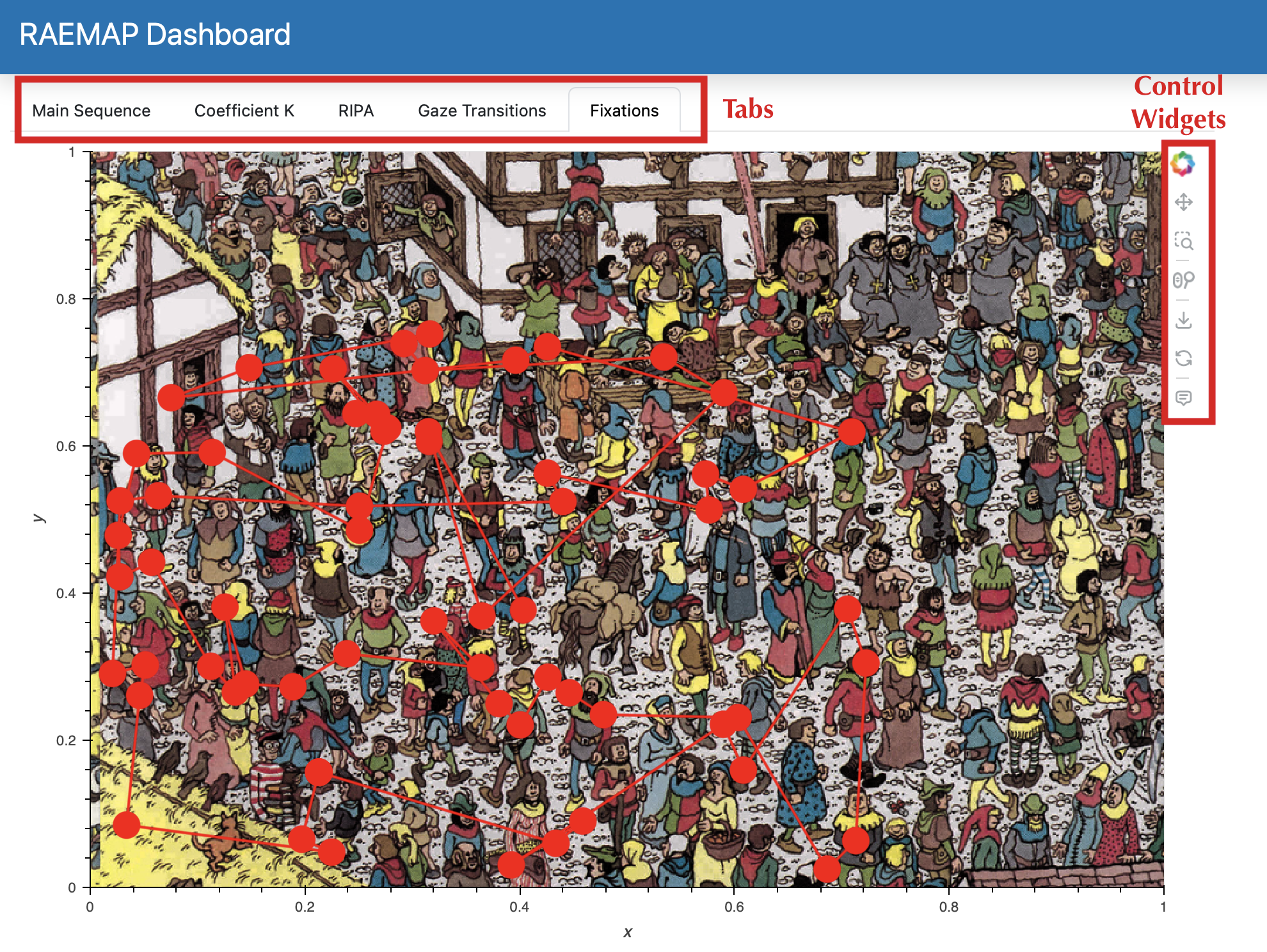}
   }
    \caption{Example Visualizations of Advanced Gaze Measures Included in the Proposed Dashboard and Key Components.
    }
    \label{fig:dashboard-all}
\end{figure}

\section{Related Work}

Eye movement recording, including pupil diameter, has been extensively used in human-computer interaction and offers the possibility of understanding how information is perceived and processed by humans \cite{mahanama2022eye}.  
Most existing software and libraries for eye-tracking analysis are hardware-specific software that comes bundled with an eye-tracking device.
They must be purchased in order to access the advanced functionalities.
Tobii Pro Lab~\cite{tobiiTrackingSoftware} and SR Research~\cite{srresearchDataViewer} are examples of hardware-specific software that comes bundled with an eye-tracking device.
Unlike vendor-specific software, iMotions~\cite{imotions} is an independent, proprietary data collection and analysis suite that is not tied to a specific vendor or device.
Also, there exist, open-source software for eye-tracking research, such as PyGaze~\cite{dalmaijer2014pygaze}, EyetrackingR~\cite{dink2015eyetrackingr}, EMA Toolbox~\cite{gibaldi2021saccade}, Open Gaze and Mouse Analyzer (OGAMA)~\cite{vosskuhler2008ogama}, PyTrack~\cite{ghose2020pytrack}, and Gaze Analytics Pipeline~\cite{duchowski2017gaze},
and research level multi-user eye tracking analysis software, such as 
DisETrac dashboard~\cite{mahanamadisetrac,abeysinghe2023disetrac,abeysinghe2024disetrac}.

Almost all of the existing eye-tracking analysis tools are capable of generating a variety of plots, including heat map generation, fixation plot generation, Areas of Interest (AOI) analysis, dynamic viewing of pupil size/gaze position, and main sequence plots.
However, these tools are intended to be used after the collected eye-tracking data is available, thus generating the aforementioned plots at the end of the eye-tracking experiment.
Moreover, except for the Gaze Analytics Pipeline~\cite{duchowski2017gaze}, and DisETrac dashboard~\cite{mahanamadisetrac,abeysinghe2023disetrac,abeysinghe2024disetrac}, none of the aforementioned tools support advanced gaze measures calculations such as ambient/focal attention coefficient $\mathcal{K}$, gaze transition entropy, or cognitive load measurements, utilizing the eye-tracking data.
Though Gaze Analytics Pipeline~\cite{duchowski2017gaze} is capable of generating advanced gaze measures and visualizations of them, similar to the other tools, it is also intended to be used after the initial process of recording eye movement data.
Alternatively, DisETrac dashboard~\cite{mahanama2020gaze, mahanamadisetrac,abeysinghe2023disetrac,abeysinghe2024disetrac} provides visualizations of traditional and advanced gaze measures
in a distributed eye tracking setting with multiple users.
However, it 
is designed to
generate visualizations for dyads, and not of a single participant,
and it does not provide visualizations of the main sequence relationship, gaze transition matrices, or the relationship between ambient/focal attention coefficient $\mathcal{K}$ and pupil diameter.

Real-Time Advanced Eye Movements Analysis Pipeline (RAEMAP)
~\cite{jayawardena2022introducing,jayawardena2020raemap}
is employed for real-time processing of advanced gaze measures.
Real-time generation of advanced gaze measures makes it possible to visualize how eye gaze measures change during the course of a single scan path of a participant during task completion.
This provides calculation of positional gaze measures such as numbers of fixations, fixation duration, average fixation duration, fixation standard deviation, pupil diameter of both eyes, percentage change of pupil diameter, maximum and minimum saccade amplitude, average saccade amplitude, and standard deviation of saccade amplitude, and main sequence analysis.

\section{Methodology}

RAEMAP~\cite{jayawardena2022introducing} integrated with StreamingHub~\cite{jayawardana2022streaminghub} provides the capability to re-stream data, mimicking the real-time simulation in the absence of real-time data. 
We utilize 
this set up to process input streams to extract gaze and pupil information and classify raw gaze points into fixations and saccades.
Upon doing so, it generates advanced gaze measures including ambient/focal attention coefficient $\mathcal{K}$~\cite{krejtz2016discerning}, an eye tracked measure of cognitive load, Real-time Index of Pupillary Activity~\cite{jayawardena2022toward},  and gaze transition matrices~\cite{krejtz2015-gaze-transition-entropy} which shows the distribution of attention over AOIs for each scan path. 
Finally, it outputs the computed advanced gaze measures as data streams.

\subsection{Advanced Gaze Measures}

\textbf{Ambient/focal attention coefficient $\mathcal{K}$}~\cite{krejtz2016discerning} is an indicator of visual search behavior. Instead of using global statistical information as in $\mathcal{K}$, 
RAEMAP~\cite{jayawardena2022introducing}
currently utilizes statistical information per subject when calculating ambient/focal attention coefficient $\mathcal{K}$. Equation~\ref{eq:window-k} shows how ambient/focal attention coefficient $\mathcal{K}$ is calculated for each fixation detected~\cite{krejtz2016discerning}.

\begin{equation}
    \mathcal{K}_i = \dfrac{d_i - \bar{d}}{\sigma_{d}} - \dfrac{a_{i+1} - \bar{a}}{\sigma_{a}}
    \label{eq:window-k}
\end{equation}

Here, $d_{i} $ is the fixation duration, and $a_{i+1} $ is the saccade amplitude. $\bar{d}_{d}, \sigma_{d}, \bar{a}_{a} \sigma_{a}$ represent the mean and standard deviation of fixation duration and saccade amplitude, of the corresponding subject.

\textbf{RIPA}~\cite{jayawardena2022toward} is an eye-tracked measure of cognitive load. It consists of short Savtizky-Golay smoothing and differentiating filters that require only a few pupillary data samples to perform the calculation. RIPA is designed to indicate a higher cognitive load if its values fall within 0.5-1.0, whereas, a lower cognitive load if its values fall within 0.0-0.5~\cite{jayawardena2022toward}.

\textbf{Gaze Transition Matrices}~\cite{krejtz2015gaze} indicate the probability of transition of gaze between two AOIs. Each cell of a gaze transition matrix corresponds to the probability calculated using the number of transitions that occurred from the source AOI to the destination AOI. 
When AOIs are considered, gaze transition matrices indicate the scan behavior of individuals with respect to the AOIs.

\textbf{Main Sequence Relationships}~\cite{bahill1975main} are plots that show the relationship between (1) the peak velocity (on the vertical axis) and amplitude (on the horizontal axis) and (2) the saccade duration (on the vertical axis) and amplitude (on the horizontal axis) of saccadic eye movements.
According to the main sequence relationships, the duration of human saccadic eye movements is related in a nonlinear manner to the saccadic amplitude, and the peak velocity is similarly related in a quasi-linear manner to saccadic amplitude, where it reaches a soft saturation limit and thereafter does not increase as rapidly.
Consequently, we hypothesize that any differences encountered in main sequence relationships could lead to the conclusion that the saccade is not normal.

\begin{figure*}[h]
    \centering
  \subfloat{
       \frame{\includegraphics[width=0.32\textwidth]{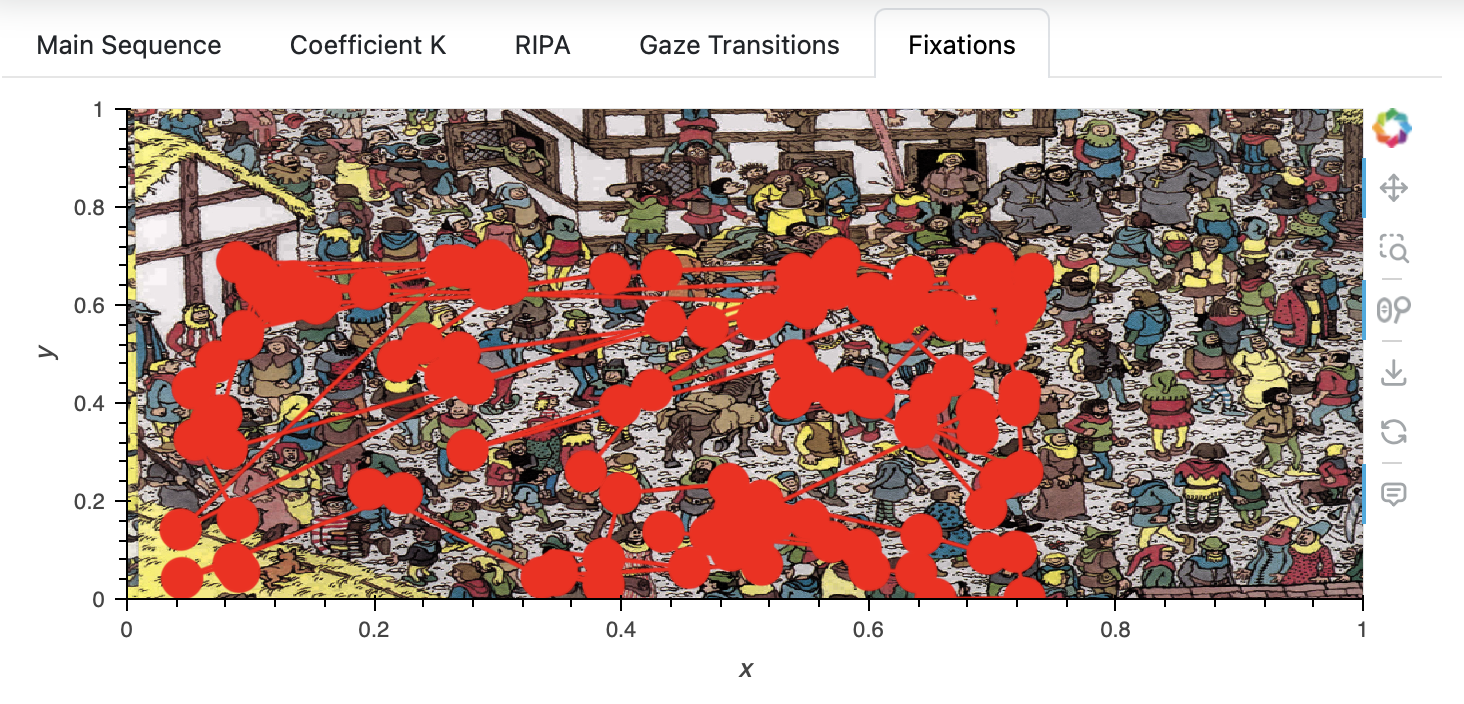}}}
  \subfloat{
        \frame{\includegraphics[width=0.32\textwidth]{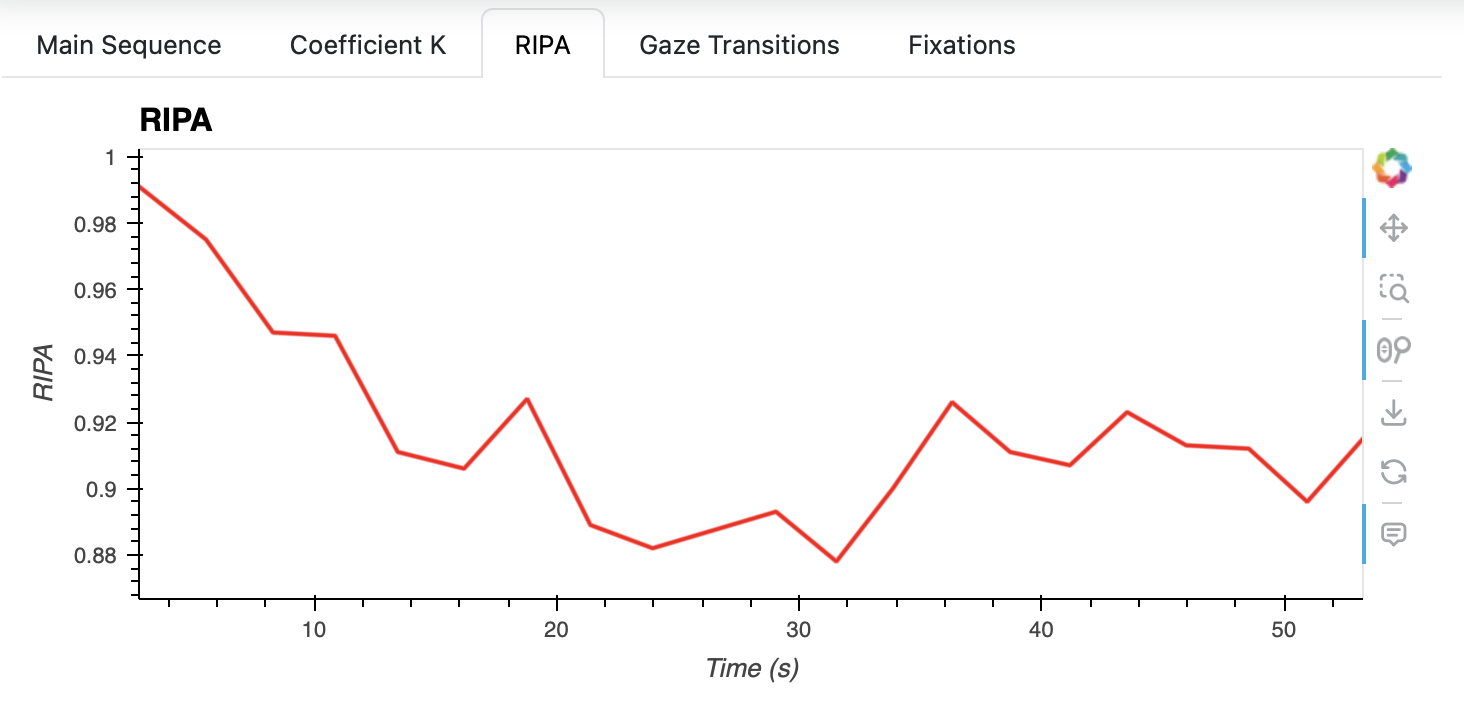}}}
  \subfloat{
        \frame{\includegraphics[width=0.3082\textwidth]{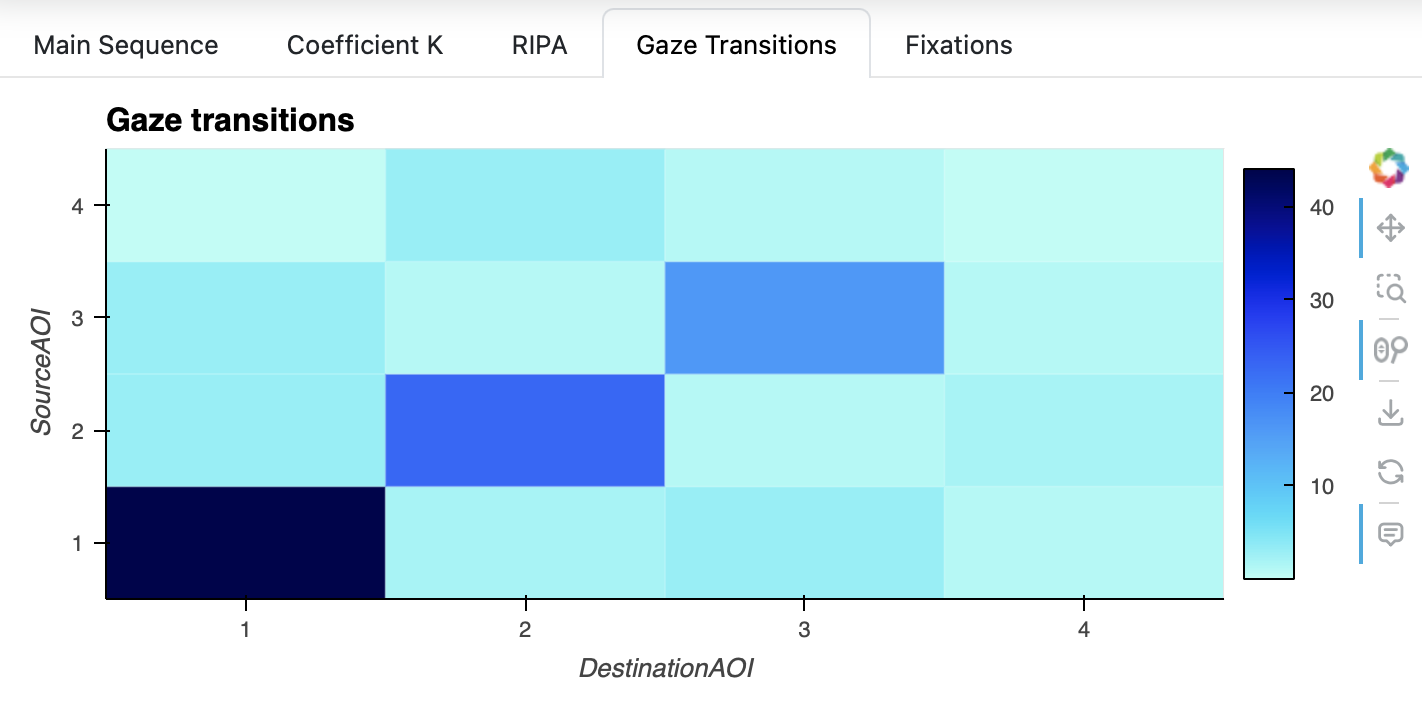}}}
  \caption{Visual Scanning - (1) Fixations Overlayed on ``Where’s Waldo'' Scene, (2) RIPA, a Measure of Cognitive Load, (3) Gaze Transitions Matrix.}
  \label{fig:vs-fxtn-ripa} 
\end{figure*}

\subsection{Advanced Gaze Analytics Dashboard}

In our setup, we acquire fixations and saccade information (including fixation centroid, fixation duration, saccade amplitude, saccade peak velocity, saccade average velocity, saccade duration) along with pupillary information, and advanced gaze measures (RIPA, ambient/focal coefficient $\mathcal{K}$, and gaze transitions between AOIs) for each subject,
by subscribing to the real-time data streams of
RAEMAP.
For each subscribed data stream, we then, generate a dynamic data frame that supports hvPlot~\cite{yang2022holoviz} to generate the visualizations.
As data is received, we emit the available data points to the corresponding stream data frame to update the visualizations.

Figure~\ref{fig:dashboard-all} shows example visualizations of our proposed dashboard along with the key components.
Our dashboard provides more interactive functionalities to monitor, analyze, and control the gaze measure visualizations. The dashboard has three main components.

\begin{enumerate}
  \item \textbf{Tabs:} enables switching between the views of different gaze measures.
  \item \textbf{Plots:} are real-time visualizations of advanced gaze measures calculated during the experiment.
  \item \textbf{Control widgets:} include zoom (box and wheel), save plot, and reset plot functions.
\end{enumerate}

There are five tabs offered in the advanced gaze measures analytics dashboard. They are the main sequence analysis tab, fixations tab, coefficient $\mathcal{K}$ tab, gaze transition matrix tab, and RIPA tab. In the \textit{main sequence analysis tab}, users can visualize both main sequence relationships: saccade amplitude vs. peak velocity, and saccade amplitude vs. saccade duration (see Figure~\ref{fig:driving-mainseqtab}).
In the \textit{fixations tab} (see the first plot in Figure
~\ref{fig:vs-fxtn-ripa}), users can analyze fixations that occurred on the scan path during the experiment. In the \textit{ambient/focal attention coefficient $\mathcal{K}$ tab} (see Figure~\ref{fig:dashboard-all}), users can visualize, coefficient $\mathcal{K}$ corresponding to each fixation duration and preceding saccade amplitude. Additionally, our dashboard lets users compare how ambient/focal attention coefficient $\mathcal{K}$ changes with fixation duration, saccade amplitude, and pupil diameter. 
In the \textit{gaze transition matrix tab} (see the third plot in Figure~\ref{fig:vs-fxtn-ripa}), users can visualize how many gaze transitions have occurred from the source AOI to the destination AOI at a given time point. The gaze transition matrix is color-coded in a way that a darker color means a higher number of transitions, whereas a lighter color means a lower number of transitions between corresponding AOIs. In the \textit{RIPA tab} (see the second plot in Figure~\ref{fig:vs-fxtn-ripa}), users can analyze RIPA as it is being generated from 
RAEMAP.

\section{Evaluation}

Using two publicly available datasets, Driving Simulation Dataset ~\cite{drivingData}, and Visual Scanning Dataset ~\cite{visualScanningDataset}, we conducted a visual analysis to validate the proposed dashboard's utility in visualizing advanced gaze measures.
For this, we use two randomly selected trials representing a scenario from both datasets and generate advanced gaze measures in real time.
We employed visualizations of fixations along the scan path, changes of ambient/focal attention coefficient $\mathcal{K}$ over time, RIPA, and main sequence relationships, compared to the anticipated behavior in both scenarios as our evaluation measures.

\subsection{Datasets}

\subsubsection{Driving Simulation Dataset}

This dataset~\cite{drivingData} contains eye movement and weighted NASA-TLX score data collected from 68 subjects who completed various driving tasks with different types of distractions, along with a non-driving baseline.
This dataset has been acquired in a controlled experiment on a driving simulator. Participants drove the same highway under four different conditions: (1) no distraction, (2) cognitive distraction, (3) emotional distraction, and (4) sensorimotor distraction.

\subsubsection{Visual Scanning Dataset}

This dataset ~\cite{visualScanningDataset} contains eye movements collected from 8 subjects who completed 120 trials of viewing images, each for 45 seconds, in two viewing conditions: (1) fixation-viewing and (2) free-viewing~\cite{visualScanningDataset}.
At each trial, participants were presented with one image from four scenes: blank scene, natural scene, picture puzzle, and Where’s Waldo.
In the free-viewing conditions, the subject’s task depended on the visual scene presented. In the blank scene and natural scene conditions, subjects were instructed to explore the image at will, and the picture puzzle and where’s waldo conditions involved visual searches.
In the fixation conditions, the visual stimulus varied from trial to trial, but the subject’s task (i.e., attempted fixation) did not.
Both fixation and free-viewing conditions were identical, except for the presence/ absence of a fixation cross.

\subsection{Real-Time Simulation and Visualizations}

Both datasets had originally split up eye-tracking data related to each trial into separate files for each participant.
To simulate real-time processing, we generated advanced gaze measures using RAEMAP.
Upon generating advanced gaze measures, it outputs the generated measures into data streams.

Our dashboard is configured to acquire the advanced gaze measures from RAEMAP
by subscribing to the real-time data streams. For each subscribed data stream, we then generate the visualizations, as data being received.
Our dashboard features five real-time visualizations of advanced gaze measures; (1) main sequence analysis,
(2) ambient/focal attention coefficient $\mathcal{K}$, 
(3) RIPA, (4) gaze transition matrix, and (5) fixations on scanpath.\\

To validate the visualizations, we consider two scenarios from the datasets.

\subsubsection{Scene One}

In a driving simulation, the expectations of a driver in terms of eye movements are similar to those in real-world driving. 
Driver should focus primarily on the road ahead to anticipate potential hazards, obstacles, and changes in traffic conditions. 
Additionally, drivers are expected to exhibit focal attention as it is a critical aspect of their cognitive engagement. 
Focal attention refers to the driver's ability to concentrate their cognitive resources on specific elements within the simulated environment. 
These elements often include the road ahead, traffic signs, other vehicles, and potential hazards. 
Maintaining focused attention is essential for making rapid and accurate decisions, effectively scanning the environment for critical information, and responding to unexpected events. 
Therefore, we anticipate drivers to exhibit focal attention during the driving task.
Moreover, cognitive processes, such as perception, attention, memory, decision-making, and problem-solving, are continually engaged as drivers assess and respond to simulated traffic situations. 
Drivers must swiftly process visual and auditory information, maintaining situational awareness by recognizing road signs, anticipating the behavior of virtual vehicles, and reacting to sudden events.
Hence we expect the driver's cognitive load to be higher in the simulation setting as well.

\begin{figure}[h]
    \centering
  \subfloat{
       \frame{\includegraphics[width=0.24\textwidth]{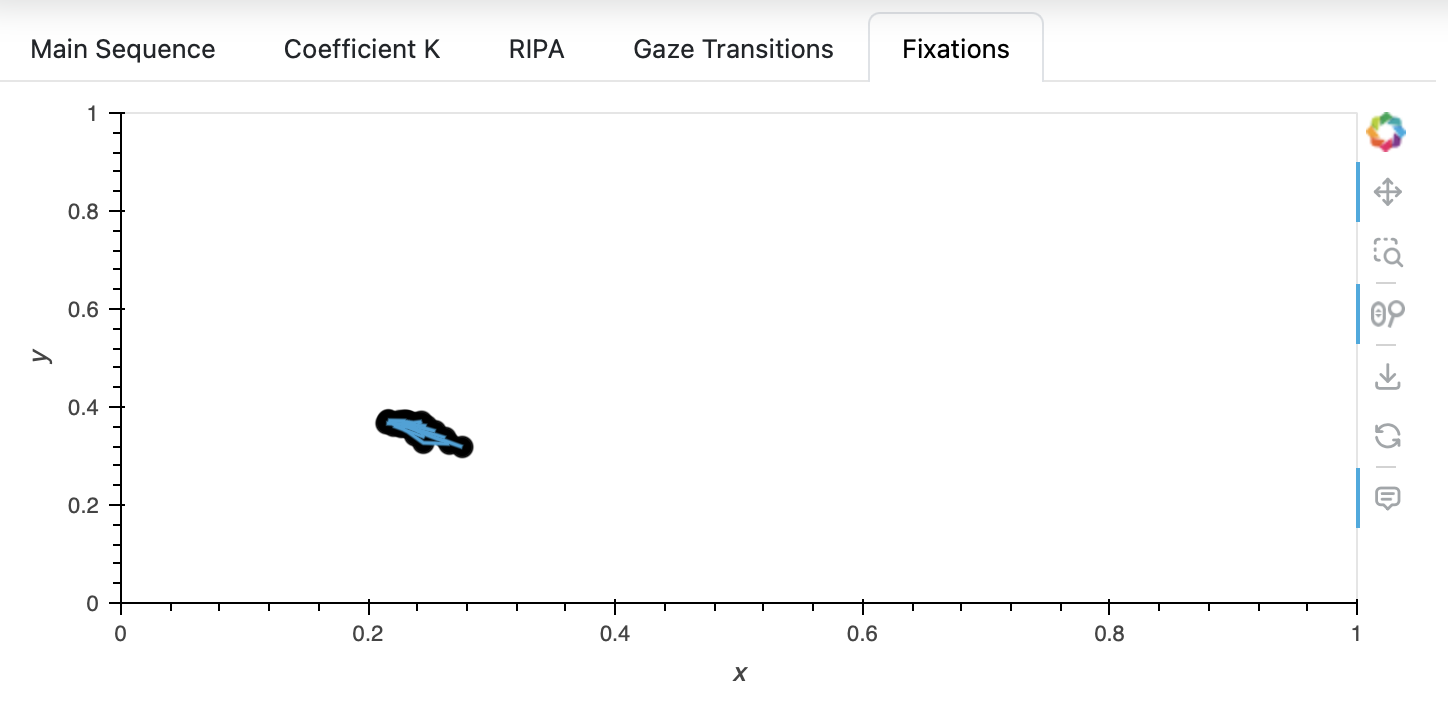}}}
  \subfloat{
        \frame{\includegraphics[width=0.24\textwidth]{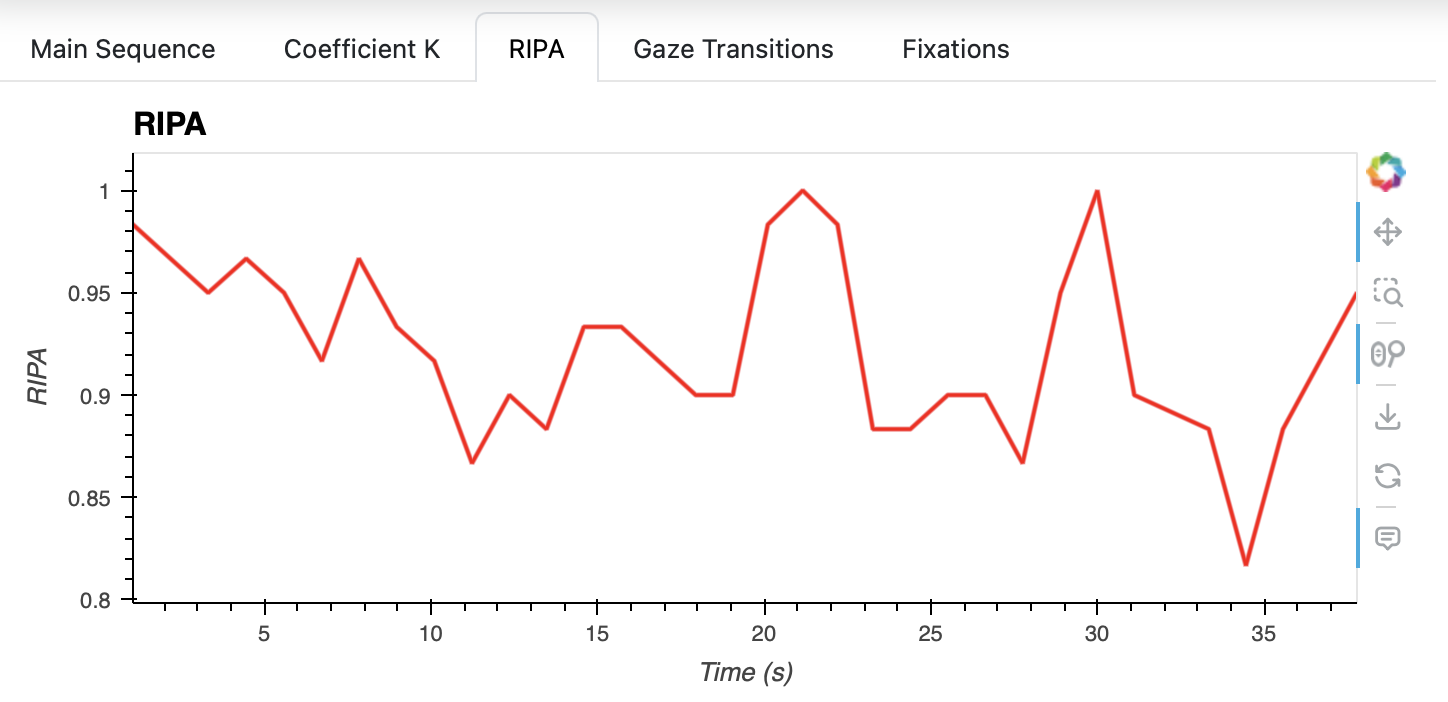}}}
  \caption{Driving Simulation - \textbf{Left: }Fixations along the Scan Path,\\ \textbf{Right:} RIPA, Eye Tracked Measure of Cognitive Load.}
  \label{fig:driving-fxtn-ripa} 
\end{figure}

We visualized the subject's fixations along the scan path while completing the driving simulation.
(see left plot in Figure~\ref{fig:driving-fxtn-ripa}).
Here, we observed that the selected subject primarily fixated on the road ahead without changing his/her gaze much.
Then, we visualized RIPA, an eye tracked measure of cognitive load (see right plot in Figure~\ref{fig:driving-fxtn-ripa}).
We observed that the subject's cognitive load varied between 0.8 and 1 on RIPA scale, indicating a higher cognitive load throughout the task.

\begin{figure}[ht]
\centering
    \frame{\includegraphics[width=0.475\textwidth]{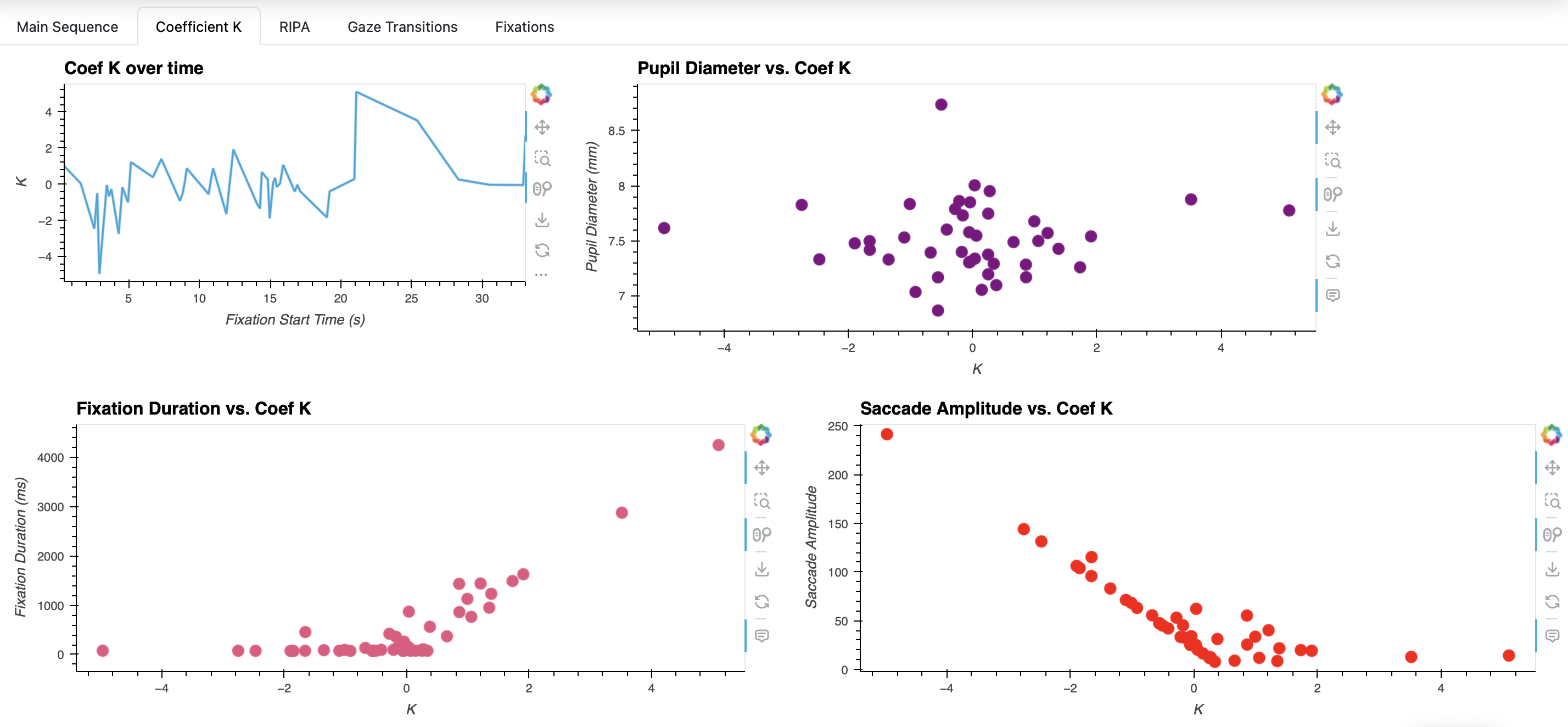}}
    \caption{Driving Simulation - Ambient/Focal Attention Coefficient $\mathcal{K}$ Over Time, and Comparisons with Pupil Diameter, Fixation Duration, and Saccade Amplitude.}
    \label{fig:driving-coefktab}
\end{figure}%

Next, we visualized the changes of ambient/focal attention coefficient $\mathcal{K}$ over time, along with comparisons between ambient/focal attention coefficient $\mathcal{K}$ and pupil diameter, fixation duration, and saccade amplitude, as illustrated in Figure~\ref{fig:driving-coefktab}.
We observed that the subject's ambient/focal attention coefficient $\mathcal{K}$ had more positive values overall.
This is an indication of a focal attention pattern.
Additionally, we observed that higher fixation durations contributed to positive ambient/focal attention coefficient $\mathcal{K}$, whereas, higher saccade amplitudes contributed to negative ambient/focal attention coefficient $\mathcal{K}$. This observation is on par with the Equation~\ref{eq:window-k}.
However, we did not observe a relationship between ambient/focal attention coefficient $\mathcal{K}$ and pupil diameter, though the subject in general had a higher pupil diameter throughout the driving task.

\begin{figure}[ht]
\centering
    \frame{\includegraphics[width=0.49\textwidth]{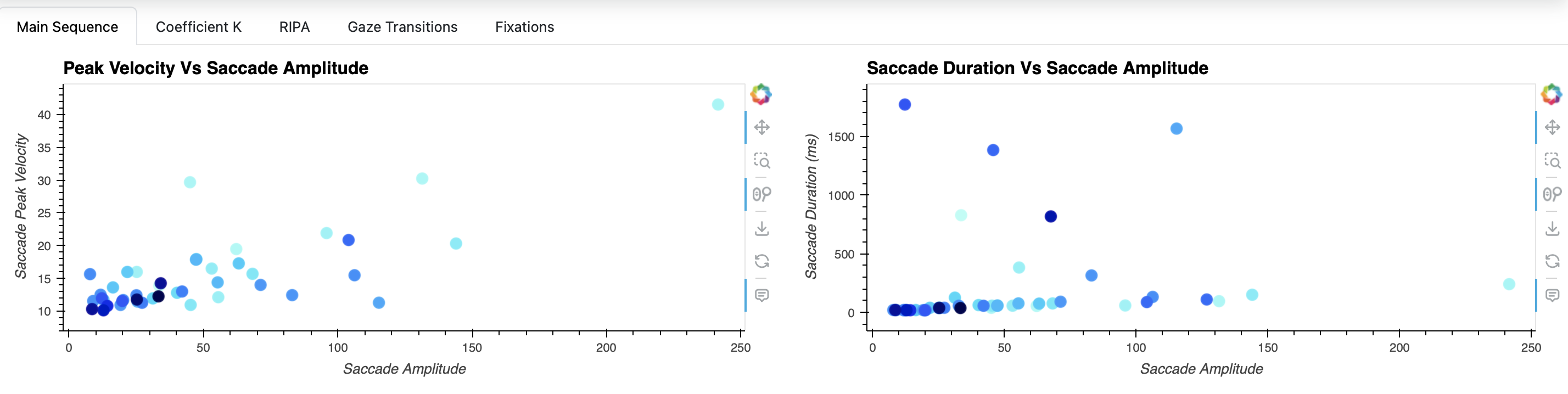}}
    \caption{Driving Simulation - Main Sequence Relationships. \textbf{Left:} Saccade Peak Velocity vs. Saccade Amplitude,
    \textbf{Right:} Saccade Duration vs. Saccade Amplitude.
    }
    \label{fig:driving-mainseqtab}
\end{figure}%

Finally, we visualized both main sequence relationships of the subject.
Figures~\ref{fig:driving-mainseqtab} show the relationships between saccade amplitude and saccade peak velocity in representative subjects during the driving task.
Here, we did not observe a quasi-linear relationship between the saccade peak velocity and the saccadic amplitude (see the left plot in Figure~\ref{fig:driving-mainseqtab}), where it reaches a soft saturation limit. It was evident that on average, saccades were not normal during driving tasks which involved different types of distractions.
However, we observed that the duration of saccadic eye movements is related in a non-linear manner to the saccadic amplitude (see the right plot in Figure~\ref{fig:driving-mainseqtab}). This observation yields that the relationship between saccade amplitude and saccade duration holds in accordance
with the main sequence relationship.

\subsubsection{Scene Two}

During visual scanning tasks, individuals typically use a combination of fixations and saccades. Fixations involve briefly pausing the eyes on a specific area of the scene to gather detailed visual information, while saccades are rapid eye movements that shift the gaze from one location to another.
In many visual search tasks, people tend to use a systematic scanning pattern. This might involve scanning from left to right or from top to bottom, or using a more complex pattern based on the task's requirements. The eye movements can be influenced by the background and context of the search task. If the target stands out from the background, it may be quickly identified. 
However, in this scenario, we selected a random subject's eye movement data collected when completing a ``where’s Waldo'' scene. Figure~\ref{fig:vs-fxtn-ripa} shows the selected ``where’s Waldo'' scene, where the \textit{Waldo} is blended into the scene, thus requiring more extensive scanning to find the target.
Due to the complexity of the scene, we expect participants to exhibit more ambient attention patterns and a higher cognitive load throughout the experiment.

We selected a random trial from the scene under the free-viewing condition, where, there was no fixation cross in the middle of the scene.
We first visualized the subject's fixations along the scan path (see the first plot in Figure~\ref{fig:vs-fxtn-ripa}) by overlaying on top of the ``Where’s Waldo'' scene. We observed that the subject exhibits a more complex scan pattern throughout the visual scan task.
Then, we visualized how RIPA changed over time (see the second plot in figure~\ref{fig:vs-fxtn-ripa}).
Here as well, we observed that the subject's cognitive load was greater than 0.8 throughout the tasks, indicating a higher cognitive load.

\begin{figure}[ht]
\centering
    \frame{\includegraphics[width=0.49\textwidth]{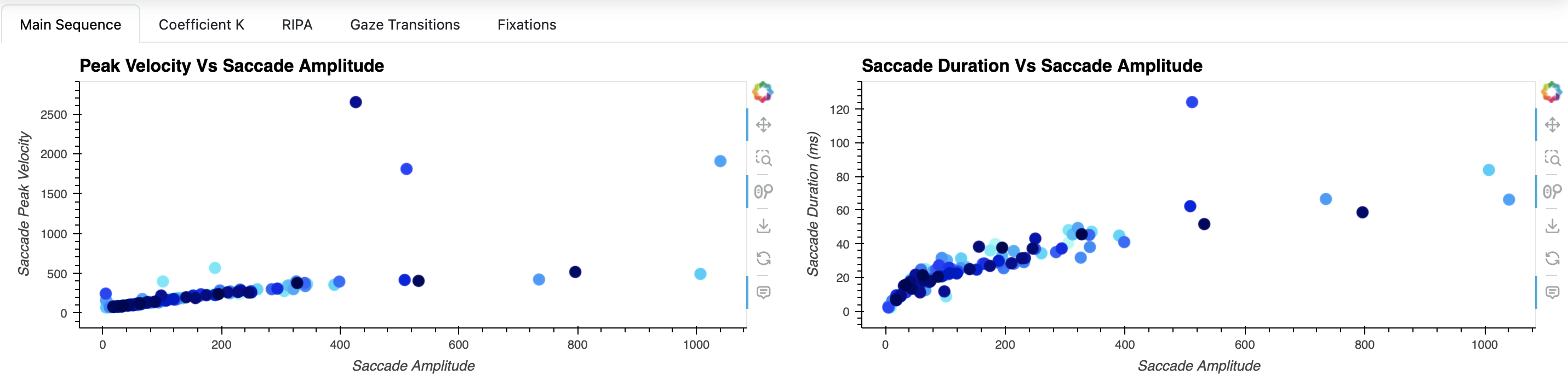}}
    \caption{Visual Scanning - Main Sequence Relationships.
    \textbf{Left:} Saccade Peak Velocity vs. Saccade Amplitude,
    \textbf{Right:} Saccade Duration vs. Saccade Amplitude.}
    \label{fig:vs-mainseqtab}
\end{figure}%

Visualizations of both main sequence relationships of the subject are shown in 
Figure~\ref{fig:vs-mainseqtab}.
Here, we observed that the saccade duration and saccade amplitude have a non-linear relationship (see the right plot in Figure~\ref{fig:vs-mainseqtab}), in accordance with the main sequence relationship.
We also observed a quasi-linear relationship between the saccade peak velocity and the saccade amplitude (see the left plot in Figure~\ref{fig:vs-mainseqtab}), showing that overall, saccades were normal during the visual scan task.

\begin{figure}[ht]
\centering
    \frame{\includegraphics[width=0.49\textwidth]{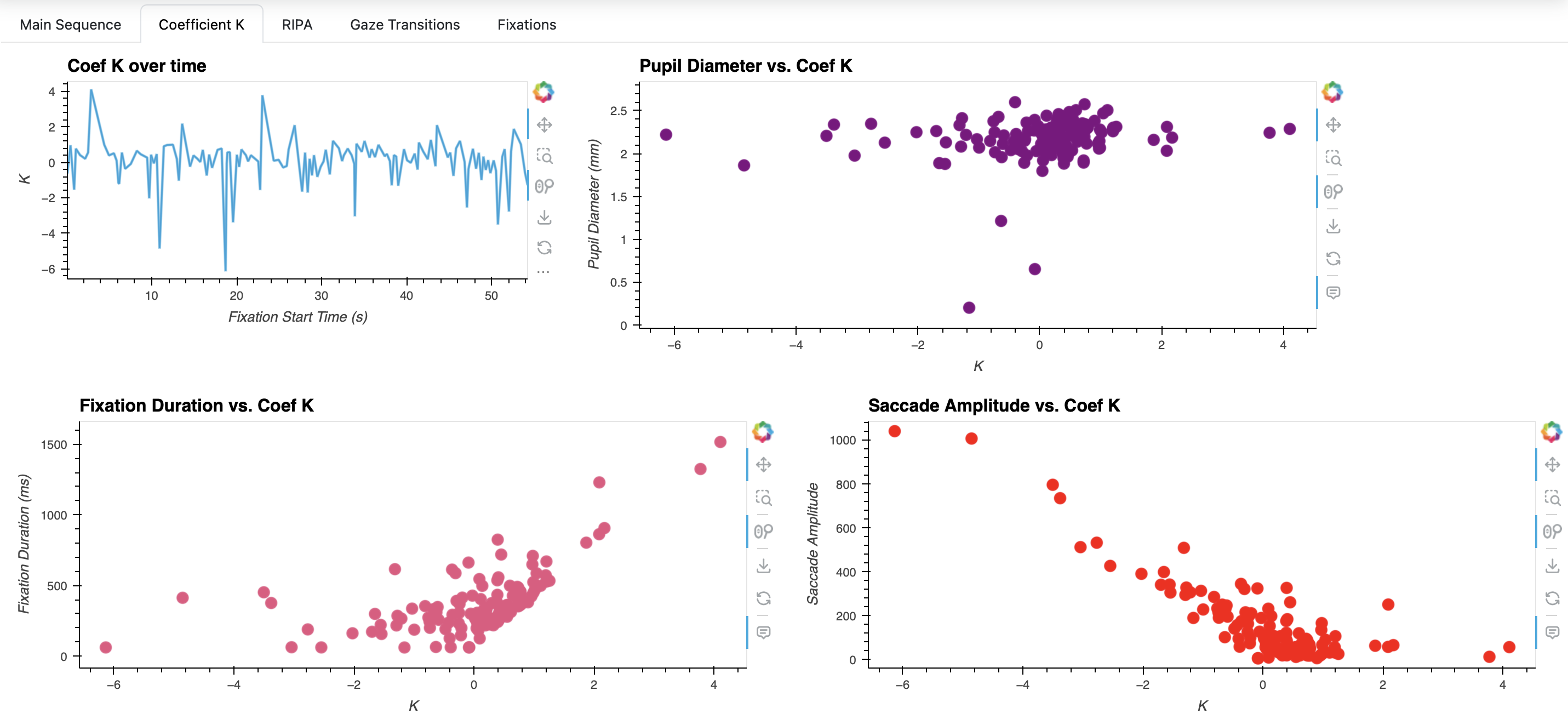}}
    \caption{Visual Scanning - Ambient/Focal Attention Coefficient $\mathcal{K}$ Over Time, and Comparisons with Pupil Diameter, Fixation Duration, and Saccade Amplitude.}
    \label{fig:vs-coefktab}
\end{figure}%

Visualizations of ambient/focal attention coefficient $\mathcal{K}$ over time (see Figure~\ref{fig:vs-coefktab}) show more negative values overall, indicating that the subject had an ambient attention pattern.
From the comparisons between ambient/focal attention coefficient $\mathcal{K}$ and pupil diameter, fixation duration, and saccade amplitude, we observed a positive linear relationship between fixation durations with ambient/focal attention coefficient $\mathcal{K}$ and a negative linear relationship between saccade amplitudes with ambient/focal attention coefficient $\mathcal{K}$.

\section{Discussion}

In line with expectations, visualizations generated from the proposed real-time gaze analytics dashboard suggest that real-time eye movement analysis indeed could provide important insights into various aspects of human behavior and cognition. 
In this study, we analyzed two distinct scenarios, where we evaluate the usability of the proposed dashboard. We monitored a driver's attention and alertness in a simulated real-time scenario, and we analyzed the visual scan behavior of a person under free viewing of a complex image followed by a search task.
In contrast to the existing eye tracking analysis tools which allow calculation and visualization of eye-tracking data upon the initial process of recording eye movement data is completed, our proposed dashboard, integrated with RAEMAP 
allows researchers to gain immediate insights into subjects' behavior such as cognitive processes, attention, and decision-making, while the experiment is ongoing in a simulated environment.

Through the visualizations, we observed a greater RIPA response during the simulated driving task as well as the complex visual search task. 
Also in line with predictions on the dynamics ambient/focal attention coefficient $\mathcal{K}$, we observed a larger number of positive coefficient $\mathcal{K}$ values in the simulated driving task, indicating a focal attention behavior, and a larger number of negative coefficient K values in the complex visual search task, indicating an ambient attention behavior.
Further studies conducted on different scenarios are required to evaluate the utility of visualizations.

While presenting promising features and insights, the suggested advanced gaze analytics dashboard does come with certain limitations.
The absence of a user evaluation is one such limitation that leaves unanswered questions regarding the practical usability and user-friendliness of the dashboard in real-world applications. 
Furthermore, the scope of the dashboard's capabilities may not encompass all potential eye movement measures and nuances that researchers might require in their work. It is essential to recognize that the effectiveness of the dashboard could vary depending on the specific research context and user requirements. 
These limitations underscore the need for continued refinement and validation as the dashboard is integrated into various research and practical settings.

Future directions of the proposed advanced gaze analytics dashboard involve refining its capabilities, expanding its applications, and ensuring it meets the evolving needs of researchers and professionals in various domains.
In the future, we plan on conducting a comprehensive user evaluation of the proposed dashboard to gather user feedback, to refine the dashboard's user interface, functionality, and accessibility.
Additionally, we also plan on enhancing the dashboard's flexibility to allow researchers to customize the measures they wish to visualize. 
Another promising future direction of the proposed gaze analytics dashboard is employing machine learning algorithms to analyze eye movement data in conjunction with other data sources. This could enable predictive analytics, aiding in areas like diagnosing cognitive disorders or predicting user behavior in user interface design.

\section{Conclusions}

In this study, we presented an eye movements analytics dashboard that enables visualizations of advanced gaze measures.
This dashboard provides visualizations of RIPA, ambient/focal attention coefficient $\mathcal{K}$, gaze transitions, and other eye movement data analysis. By using two publicly available eye-tracking datasets, we showcased the proposed dashboard's utility in visualizing advanced eye movement measures generated using multiple data sources. In the future, we plan to improve the proposed dashboard and conduct a user study to assess the effectiveness of our dashboard by recruiting experts in UI/UX design and the eye-tracking domain.

\section*{Acknowledgment}

This work was supported in part by the National Science Foundation \# 
2045523.


\bibliographystyle{ieeetr}
\bibliography{bibliography}

\begin{thebibliography}{10}

\bibitem{just1980theory}
M.~A. Just and P.~A. Carpenter, ``A theory of reading: from eye fixations to comprehension.,'' {\em Psychological review}, vol.~87, no.~4, p.~329, 1980.

\bibitem{popa2015reading}
L.~Popa, O.~Selejan, A.~Scott, D.~F. Mure{\c{s}}anu, M.~Balea, and A.~Rafila, ``Reading beyond the glance: eye tracking in neurosciences,'' {\em Neurological Sciences}, vol.~36, no.~5, pp.~683--688, 2015.

\bibitem{jayawardena2020pilot}
G.~Jayawardena, A.~Michalek, A.~Duchowski, and S.~Jayarathna, ``Pilot study of audiovisual speech-in-noise (sin) performance of young adults with adhd,'' in {\em ACM Symposium on Eye Tracking Research and Applications}, (Stuttgart, Germany), pp.~1--5, ACM, 2020.

\bibitem{ahlstrom2021eye}
C.~Ahlstr{\"o}m, K.~Kircher, M.~Nystr{\"o}m, and B.~Wolfe, ``Eye tracking in driver attention research—how gaze data interpretations influence what we learn,'' {\em Frontiers in neuroergonomics}, vol.~2, p.~778043, 2021.

\bibitem{krejtz2016discerning}
K.~Krejtz, A.~Duchowski, I.~Krejtz, A.~Szarkowska, and A.~Kopacz, ``Discerning ambient/focal attention with coefficient k,'' {\em ACM Transactions on Applied Perception (TAP)}, vol.~13, no.~3, pp.~1--20, 2016.

\bibitem{jayawardena2022toward}
G.~Jayawardena, Y.~Jayawardana, S.~Jayarathna, J.~H{\"o}gstr{\"o}m, T.~Papa, D.~Akkil, A.~T. Duchowski, V.~Peysakhovich, I.~Krejtz, N.~Gehrer, {\em et~al.}, ``Toward a real-time index of pupillary activity as an indicator of cognitive load,'' {\em Procedia Computer Science}, vol.~207, pp.~1331--1340, 2022.

\bibitem{krejtz2015-gaze-transition-entropy}
K.~Krejtz, A.~Duchowski, T.~Szmidt, I.~Krejtz, F.~Gonz{\'a}lez~Perilli, A.~Pires, A.~Vilaro, and N.~Villalobos, ``Gaze transition entropy,'' {\em ACM Transactions on Applied Perception (TAP)}, vol.~13, no.~1, pp.~1--20, 2015.

\bibitem{mahanama2022eye}
B.~Mahanama, Y.~Jayawardana, S.~Rengarajan, G.~Jayawardena, L.~Chukoskie, J.~Snider, and S.~Jayarathna, ``Eye movement and pupil measures: A review,'' {\em frontiers in Computer Science}, vol.~3, p.~733531, 2022.

\bibitem{tobiiTrackingSoftware}
``{E}ye tracking software for behavior research --- tobii.com.'' \url{https://www.tobii.com/products/software/behavior-research-software/tobii-pro-lab}.
\newblock [Accessed 12-04-2024].

\bibitem{srresearchDataViewer}
``{D}ata {V}iewer --- sr-research.com.'' \url{https://www.sr-research.com/data-viewer/}.
\newblock [Accessed 12-04-2024].

\bibitem{imotions}
``i{M}otions | {P}owering {H}uman {I}nsights - {B}iometric {R}esearch --- imotions.com.'' \url{https://imotions.com}.
\newblock [Accessed 12-04-2024].

\bibitem{dalmaijer2014pygaze}
E.~S. Dalmaijer, S.~Math{\^o}t, and S.~Van~der Stigchel, ``Pygaze: An open-source, cross-platform toolbox for minimal-effort programming of eyetracking experiments,'' {\em Behavior research methods}, vol.~46, no.~4, pp.~913--921, 2014.

\bibitem{dink2015eyetrackingr}
J.~W. Dink and B.~Ferguson, ``eyetrackingr: An r library for eye-tracking data analysis,'' 2015.

\bibitem{gibaldi2021saccade}
A.~Gibaldi and S.~P. Sabatini, ``The saccade main sequence revised: A fast and repeatable tool for oculomotor analysis,'' {\em Behavior Research Methods}, vol.~53, no.~1, pp.~167--187, 2021.

\bibitem{vosskuhler2008ogama}
A.~Vo{\ss}k{\"u}hler, V.~Nordmeier, L.~Kuchinke, and A.~M. Jacobs, ``Ogama (open gaze and mouse analyzer): open-source software designed to analyze eye and mouse movements in slideshow study designs,'' {\em Behavior research methods}, vol.~40, no.~4, pp.~1150--1162, 2008.

\bibitem{ghose2020pytrack}
U.~Ghose, A.~A. Srinivasan, W.~P. Boyce, H.~Xu, and E.~S. Chng, ``Pytrack: An end-to-end analysis toolkit for eye tracking,'' {\em Behavior research methods}, vol.~52, no.~6, pp.~2588--2603, 2020.

\bibitem{duchowski2017gaze}
A.~T. Duchowski, ``The gaze analytics pipeline,'' in {\em Eye Tracking Methodology}, pp.~175--191, New York, NY: Springer, 2017.

\bibitem{mahanamadisetrac}
B.~Mahanama, M.~Sunkara, V.~Ashok, and S.~Jayarathna, ``Disetrac: Distributed eye-tracking for online collaboration,'' in {\em Proceedings of the 2023 Conference on Human Information Interaction and Retrieval}, CHIIR '23, (New York, NY, USA), p.~427–431, Association for Computing Machinery, 2023.

\bibitem{abeysinghe2023disetrac}
Y.~Abeysinghe, B.~Mahanama, G.~Jayawardena, M.~Sunkara, V.~Ashok, and S.~Jayarathna, ``Gaze analytics dashboard for distributed eye tracking,'' in {\em 2023 IEEE 24th International Conference on Information Reuse and Integration for Data Science (IRI)}, pp.~140--145, 2023.

\bibitem{abeysinghe2024disetrac}
Y.~Abeysinghe, B.~Mahanama, G.~Jayawardena, Y.~Jayawardana, M.~Sunkara, A.~T. Duchowski, V.~Ashok, and S.~Jayarathna, ``A-disetrac advanced analytic dashboard for distributed eye tracking,'' {\em International Journal of Multimedia Data Engineering and Management (IJMDEM)}, vol.~15, no.~1, pp.~1--20, 2024.

\bibitem{mahanama2020gaze}
B.~Mahanama, Y.~Jayawardana, and S.~Jayarathna, ``Gaze-net: appearance-based gaze estimation using capsule networks,'' in {\em Proceedings of the 11th Augmented Human International Conference}, pp.~1--4, 2020.

\bibitem{jayawardena2022introducing}
G.~Jayawardena, ``Introducing a real-time advanced eye movements analysis pipeline,'' in {\em 2022 Symposium on Eye Tracking Research and Applications}, pp.~1--2, 2022.

\bibitem{jayawardena2020raemap}
G.~Jayawardena, ``Raemap: Real-time advanced eye movements analysis pipeline,'' in {\em Symposium on Eye Tracking Research and Applications 2020}, (Stuttgart, Germany), ACM, 2020.

\bibitem{jayawardana2022streaminghub}
Y.~Jayawardana, V.~G. Ashok, and S.~Jayarathna, ``Streaminghub: Interactive stream analysis workflows,'' in {\em Proceedings of the 22nd ACM/IEEE Joint Conference on Digital Libraries}, JCDL '22, (New York, NY, USA), Association for Computing Machinery, 2022.

\bibitem{krejtz2015gaze}
K.~Krejtz, A.~Duchowski, T.~Szmidt, I.~Krejtz, F.~Gonz{\'a}lez~Perilli, A.~Pires, A.~Vilaro, and N.~Villalobos, ``Gaze transition entropy,'' {\em ACM Transactions on Applied Perception (TAP)}, vol.~13, no.~1, p.~4, 2015.

\bibitem{bahill1975main}
A.~T. Bahill, M.~R. Clark, and L.~Stark, ``The main sequence, a tool for studying human eye movements,'' {\em Mathematical biosciences}, vol.~24, no.~3-4, pp.~191--204, 1975.

\bibitem{yang2022holoviz}
S.~Yang, M.~S. Madsen, and J.~A. Bednar, ``Holoviz: Visualization and interactive dashboards in python,'' in {\em Proceedings of the 28th ACM SIGKDD Conference on Knowledge Discovery and Data Mining}, pp.~4846--4847, 2022.

\bibitem{drivingData}
S.~Taamneh, P.~Tsiamyrtzis, M.~Dcosta, P.~Buddharaju, A.~Khatri, M.~Manser, T.~Ferris, R.~Wunderlich, and I.~Pavlidis, ``A multimodal dataset for various forms of distracted driving,'' {\em Scientific Data}, 2017.

\bibitem{visualScanningDataset}
M.~B. McCamy, J.~Otero-Millan, L.~L. Di~Stasi, S.~L. Macknik, and S.~Martinez-Conde, ``Highly informative natural scene regions increase microsaccade production during visual scanning,'' {\em Journal of neuroscience}, vol.~34, no.~8, pp.~2956--2966, 2014.

\end{thebibliography}

\end{document}